\documentclass[prb,showpacs,superscriptaddress,footinbib]{revtex4}

\usepackage[latin1]{inputenc}
\usepackage{epsfig}
\usepackage{dcolumn}
\usepackage{bm}
\usepackage{amsmath}
\usepackage{amsfonts}
\usepackage{amssymb}
\usepackage{wrapfig}
\usepackage{subfigure}
\usepackage{psfrag}
\usepackage{tabularx}
\usepackage{graphicx}

\DeclareMathOperator*{\arcsinh}{arcsinh}

\begin{document}

\newcommand{\sign}{\operatorname{sign}}
\newcommand{\Ci}{\operatorname{Ci}}
\newcommand{\Si}{\operatorname{Si}}
\newcommand{\tr}{\operatorname{tr}}

\newcommand{\beq}{\begin{equation}}
\newcommand{\eeq}{\end{equation}}
\newcommand{\beqn}{\begin{eqnarray}}
\newcommand{\eeqn}{\end{eqnarray}}

\newcommand{\slp}{\raise.15ex\hbox{$/$}\kern-.57em\hbox{$ \partial $}}
\newcommand{\lnA}{\raise.15ex\hbox{$/$}\kern-.57em\hbox{$A$}}
\newcommand{\unmedio}{{\scriptstyle\frac{1}{2}}}
\newcommand{\uncuarto}{{\scriptstyle\frac{1}{4}}}

\newcommand{\trial}{_{\text{trial}}}
\newcommand{\true}{_{\text{true}}}
\newcommand{\const}{\text{const}}

\newcommand{\intp}{\int\frac{d^2p}{(2\pi)^2}\,}
\newcommand{\intx}{\int_C d^2x\,}
\newcommand{\inty}{\int_C d^2y\,}
\newcommand{\intxy}{\int_C d^2x\,d^2y\,}

\newcommand{\bP}{\bar{\Psi}}
\newcommand{\bc}{\bar{\chi}}
\newcommand{\hs}{\hspace*{0.6cm}}

\newcommand{\bra}{\left\langle}
\newcommand{\ket}{\right\rangle}
\newcommand{\bracket}{\left\langle\,\right\rangle}

\newcommand{\D}{\mbox{$\mathcal{D}$}}
\newcommand{\N}{\mbox{$\mathcal{N}$}}
\newcommand{\Lag}{\mbox{$\mathcal{L}$}}
\newcommand{\V}{\mbox{$\mathcal{V}$}}
\newcommand{\Z}{\mbox{$\mathcal{Z}$}}
\newcommand{\A}{\mbox{$\mathcal{A}$}}
\newcommand{\B}{\mbox{$\mathcal{B}$}}
\newcommand{\C}{\mbox{$\mathcal{C}$}}
\newcommand{\E}{\mbox{$\mathcal{E}$}}



\title{Transient effects in the backscattered current of a Luttinger liquid}
\author{Mariano J. Salvay}
\author{Hugo A. Aita}
\author{Carlos M. Na\'on}
\affiliation{Departamento de F\'{\i}sica, Facultad de Ciencias
Exactas, Universidad Nacional de La Plata and IFLP-CONICET, CC 67,
 1900 La Plata, Argentina.}

\begin{abstract}We study the backscattered current in a Luttinger
liquid in the presence of a point like weak impurity switched on at
finite time, taking into account finite-temperature effects.  We
show how the well-known results for a static impurity are distorted.
We derive a dimensionless parameter $\tau_{R}$ as function of the
electron-electron interaction and the temperature, such that for
$\tau_{R} < 1$ ($> 1$) the switching process is relevant
(irrelevant).  Our results suggest the possibility of determining
the value of the Luttinger parameter $K$ through time measurements
in transport experiments at fixed voltage.
\end{abstract}
 \maketitle

\section{Introduction}
Quantum transport in novel one-dimensional (1D) materials, such as
quantum wires and carbon nanotubes \cite{CNT}, is one of the most
active areas of present research in condensed matter physics. Recent
experiments have confirmed some of the striking effects that
characterize the Luttinger liquid (LL) picture of 1D nanostructures
\cite{Luttinger}, such as spin-charge separation \cite{spin-charge}
and charge fractionalization \cite{fractionalization}. One central
issue in all experiments is the determination of Luttinger exponents
that depend on the rigidity constant $K$, which parametrizes
electron-electron (e-e) interactions. These exponents have been
measured in pioneering tunneling experiments \cite{Exp-exponents}.
However, some doubts still remain in the determination of $K$ due to
the fact that dynamical Coulomb blockade also leads to power laws in
systems surrounded by ohmic devices \cite{ohmic}. Moreover, it has
been shown that environmental resistance can also contribute to the
measured value of $K$ \cite{environment}. It is then very important
to conceive alternative ways to measure $K$. To a large extent,
quantum transport is highly non trivial due to the influence of a
variety of combined effects produced by impurities, junctions,
contacts, etc. In particular, in the study of impurities in
Luttinger liquids, an observable of special interest is the
backscattered current $I_{bs}$.  For a point like static impurity,
$I_{bs}$ opposes the background current and goes as $V^{2 K - 1}$,
where $V$ is the bias voltage \cite{Kane}. The effects of finite
temperature and finite length of the quantum wire lead to
characteristic non-monotonic parameter dependencies  of the current
and yield a rich structure in the noise spectrum which depends on
$K$ \cite{Dolcini}. For a time-dependent oscillatory impurity,  the
current grows and the conductance of a one-channel quantum wire is
greater than its background value $e^{2}/h$ for strong repulsive
interaction ($K < 1/2$) \cite{Feldman}\,\cite{cheng-zhou}.  More
recently, the effect of several time-dependent impurities was
considered at zero temperature and infinite
length\cite{theories}\,\cite{makogon}. When temperature is taken
into account, for two oscillating barriers, the previous zero
temperature results are distorted; the behavior of the backscattered
current and the change in the differential conductance depend on
different regimes which can be established as function of the
frequency, the temperature and the separation between the impurity
potentials \cite{nos 2}.

One aspect that has been seldom explored in the context of the
Luttinger liquid is its response to a sudden switch of an
interaction of the system with an external field.  As an example,
the total energy density of a Tomonaga-Luttinger liquid in the
presence of a forward-scattering time-dependent impurity switched on
at finite time has been computed exactly, distinguishing two well
defined regimes in terms of the relationship between the frequency
of the perturbation and the electron energy. This study allowed to
determine a time interval in which the finite-time switching process
is relevant \cite{nos 1}.

In this paper we examine the effect of a transient process on
transport properties of a Luttinger liquid. When a local barrier,
that can be considered as a backscattering impurity, is turned on at
a finite time $t_0$, a backscattered current is produced. Let us
stress that such a barrier can be created, for instance, in a
single-walled carbon nanotube by applying a voltage to a narrow
metal gate electrode \cite{biercuk} (See Fig. 1). We obtain an
analytical expression for $I_{bs}$ as function of time. The envelope
of this function decays in a way that crucially depends on e-e
interactions, i. e. on the $K$ parameter. This allows us to find a
simple relation between the decay process and the value of $K$.
Thus, we show that $K$ could be determined by measuring time
intervals within the reach of recently developed pump-probe
techniques with femtosecond-attosecond time resolutions
\cite{femtosecondExp}. Then, in contrast to the conventional
techniques used up to now to measure $K$, based on the study of
stationary transport properties for varying voltages, we indicate an
alternative way based on the analysis of the transient current for
fixed voltage. Apart from its interest in the context of strongly
correlated electronic systems, our studies could be relevant in the
area of cold atomic gases, where quantum quenches are being
intensively investigated \cite{quenches}.

The work is organized as follows.  In Section II we present the
model and review the calculation of the backscattered current in
terms of a vacuum expectation value of exponentials of bosonic
fields. In Sections III and IV we present the results obtained at
zero and finite temperature, respectively. We define a dimensionless
relaxation parameter $\tau_{R}$ as function of the temperature and
of the electron-electron interaction.  Finally, we analyzed and
summarize our results in Sec V.

\section{The model}
\begin{figure}\begin{center}
\includegraphics{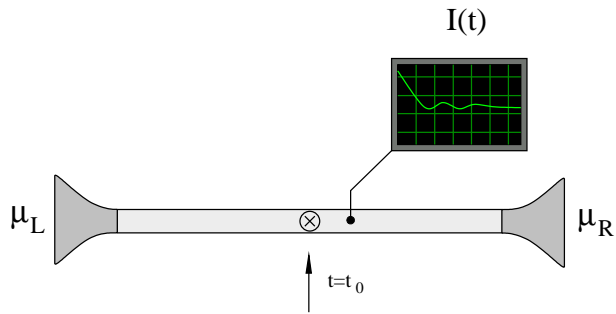}
\caption{\label{fig2ne}The figure shows a quantum wire coupled adiabatically to two reservoirs with different chemical potentials,
with a backscattering impurity switched on at time $t=t_0$. The current $I(t)$ is measured as a function of time.}
\end{center}\end{figure}

We consider a long and clean LL adiabatically coupled to two
electrodes with different chemical potentials, such that $\mu_{L}
-\mu_{R} = e V$. We will restrict our study to the case in which the
electrodes are held at the same temperature. This condition is very
important in order to apply standard bosonization techniques
\cite{standardbos}. Indeed, as was very recently explained
\cite{GGM}, standard bosonization is expected to work well only in
special situations corresponding to small deviations from
equilibrium. The non trivial ingredient of our model is a barrier of
constant height (an externally controlled impurity) that is turned
on at a finite time $t_0$, producing backscattering of incident
waves. Under these conditions we model the LL in terms of the
following Lagrangian density \cite{makogon}:

\begin{equation} L = L_{0} + L_{imp} \, ,
\end{equation} where\begin{equation} L_{0} = \frac{1}{2}\Phi(x,t)
\left(v^{2}\frac{\partial^{2}}{\partial_{x}^{2}} -
\frac{\partial^{2}}{\partial_{t}^{2}}\right)\Phi(x,t) \,
\end{equation} describes a spinless Tomonaga-Luttinger liquid with
renormalized velocity  $v$ and\begin{equation} L_{imp}= - \frac{
g_{B}}{\pi \hbar \Lambda} \delta(x - x_{0}) \Theta (t -
t_{0})\cos\left[2 k_{F}x/\hbar + 2\sqrt{\pi K v}\Phi(x, t) + e V
t/\hbar\right] \, \end{equation}represents the scattering of
spinless electrons with the external barrier at the point $x_{0}$
and switched on at the time $t_{0}$, with a coupling amplitude
$g_{b}$. $V$ is the external voltage applied to the quantum wire and
$K$ measures the strength of the electron-electron interactions. For
repulsive interactions $K < 1$, and for noninteracting electrons $K
= 1$. $\Lambda$ is a short-distance cutoff. In the above expression
we only take into account  backscattering between electrons and
impurities, because the forward scattering does not change the
transport properties studied here, at least, in the lowest-order of
the perturbative expansion in the couplings.

In the absence of the impurity, the background current is $I_{0} =
e^{2} V/h$ . When the impurity is acting the total current is $I =
I_{0} - I_{bs}$. The operator associated to the backscattered
current is defined as \cite{makogon}
\begin{equation}\widehat{I}_{bs}(t) = \frac{g_{B} e}{\pi \hbar \Lambda} \Theta (t - t_{0}) \sin[2 k_{F}x_{0}/\hbar + 2\sqrt{\pi K
v}\widehat{\Phi}(x_{0}, t) + e V t/\hbar] \, .\end{equation}

The backscattered current at any time $t$ is given by\beq I_{bs}(t)
= \langle 0|S(- \infty ; t)\widehat{I}_{bs}(t)S(t ; - \infty) | 0
\rangle \, ,  \label{uno} \eeq where $ \langle 0|$ denotes the
initial state and $S$ is the scattering matrix, which to the lowest
order in the coupling $g_{B}$ is given by\beq S(t ; - \infty) = 1 -
i \int^{\infty}_{-\infty} d x \int^{t}_{- \infty} L_{imp} (t') d t'
\,\label{dos} .\eeq

When one inserts (\ref{dos}) in (\ref{uno}) one finds several terms
of the form

\beq A_{\alpha,\beta}=\langle 0|\exp[ 2 i \alpha \sqrt{\pi K v}
\widehat{\Phi}(x, t')]\exp[- 2 i \beta \sqrt{\pi K v}
\widehat{\Phi}(x, t)] |0 \rangle, \eeq with $\alpha,\beta=\pm 1$.
This kind of vacuum expectation values (v.e.v.) of vertex operators
have been computed many times in the literature. It is well-known
that $A_{\alpha,-\alpha}=0$. Thus, the building block of our
computation is $A_{\alpha,\alpha}$. Let us sketch the calculational
procedure for $\alpha=1$. Using Baker-Haussdorff formula and taking
into account that the commutator of the fields is a c-number, we can
write:

\beq A_{1,1}=\langle 0|\exp\big[ 2 i \sqrt{\pi K v}
(\widehat{\Phi}(x, t')-\widehat{\Phi}(x, t))\big]| 0 \rangle
\,\exp\big[2 \pi K v [\widehat{\Phi}(x, t'),\widehat{\Phi}(x,
t)]\big]. \eeq

At this point we observe that the v.e.v. of the exponential in the
first factor above, can be written as the exponential of a v.e.v.:

\beq \langle 0|\exp\big[ 2 i \sqrt{\pi K v} (\widehat{\Phi}(x,
t')-\widehat{\Phi}(x, t))\big]| 0 \rangle = \exp\big[-2 \pi K v \,
\langle 0|(\widehat{\Phi}(x, t')-\widehat{\Phi}(x, t))^2 | 0 \rangle
\big].\eeq

Now, in order to explicitly evaluate the previous expressions we
need Keldysh \cite{Keldysh} lesser function $G^{<}$ given by

\beq \langle 0|\widehat{\Phi}(x, t')\,\widehat{\Phi}(x, t))\big)| 0
\rangle = i
G^{<}(x,t;x,t')=\frac{1}{2\pi}\,\int\,dp\,d\omega\,e^{-i\omega(t-t')}\,\big(\theta(-\omega)+n_B(|\omega|)\big)\,\delta(\omega^2-v^2p^2),\eeq
where $n_B(|\omega|)=(e^{\frac{\hbar|\omega|}{k_B T}}-1)^{-1}$.
Putting all this together we finally obtain

\begin{equation} \langle 0|\exp[i 2 \sqrt{\pi K v}
\widehat{\Phi}(x, t')]\exp[- i 2 \sqrt{\pi K v} \widehat{\Phi}(x,
t)] 0 \rangle  =  \frac{ (\Lambda \pi k_{b} T /\hbar v)^{2 K} \exp[i
\pi K sgn [t - t']]}{\left|\sinh[\pi k_{b} T (t -
t')/\hbar]\right|^{2 K}} \, , \label{tres}
\end{equation}where $sgn$ is the Sign function.

\section{Results at zero temperature}
\begin{figure}\begin{center}
\psfrag{I}[c][t]{$I(t)$}
\includegraphics{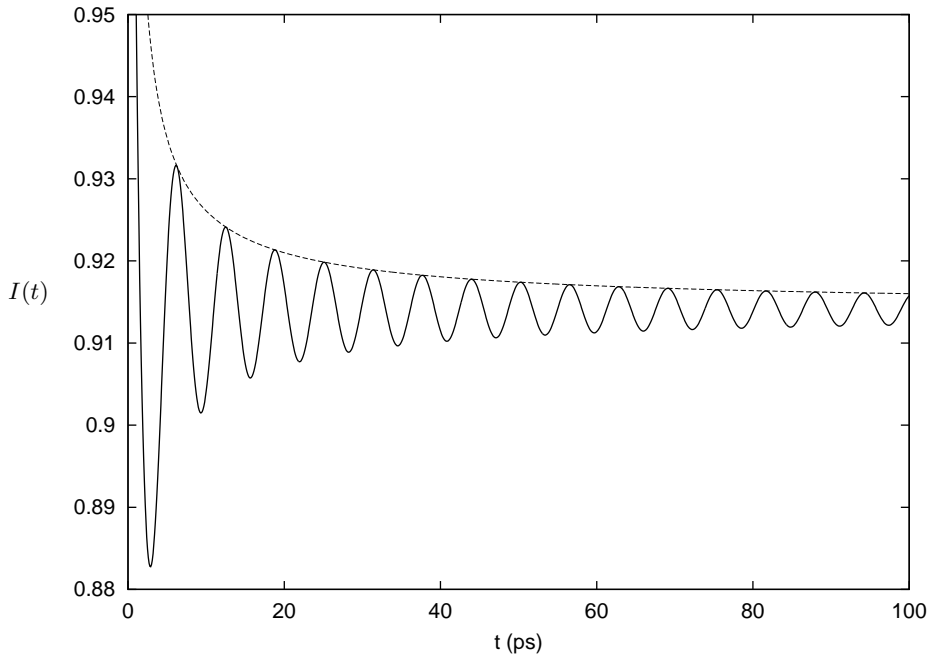}
 \caption{\label{fig3}Example of an experimental plot of $I$ as function $t$, leading to the determination of
$K$. We have set $\frac{e V}{\hbar} = 10^{12}(1/seg)$ (corresponding
to $V\simeq mV$), $\frac{g_{B}}{\hbar v}= (0.1)^{1/2}$,
$\frac{\Lambda e V}{ \hbar v}= 1$. $I$ is measured in units of
$\frac{e^2 V}{h}$. For this particular case, using the envelope
(dashed line) one obtains $K=0.4$. }
\end{center}\end{figure}
First of all, we compute (\ref{uno}) at zero temperature and with
the impurity switched on at time $- \infty$ :

\begin{equation} I_{bs}(t, -\infty) = \frac{g_{B}^{2} e  \Lambda ^{2 K - 2}}{2 \pi \hbar^{2} v ^{2 K}\Gamma[2 K]}\left|\frac{e V}{\hbar}\right|^{2 K -1} sgn [V].\label{pu}\end{equation}
This corresponds to the well-known case of a static
impurity\cite{Kane}, where the backscattered current goes as $V^{2 K
- 1}$. We note that for $K < 1/2$, the backscattered current becomes
large when $V$ decreases. Hence, the perturbative expansion in
powers of $g_{B}$ breaks down when $V \rightarrow 0$.  Using a
scaling analysis we can estimate that this expansion is valid when
$\frac{g_{B}}{\hbar v} (\frac{\Lambda e V}{ \hbar v})^{K - 1} \ll
1$. We emphasize that expression (\ref{pu}) does not include the
case $V = 0$, where the current is zero too.  All these statements
imply that the current must be a nonmonotonic function of $V$.  In
order to determine this function one has to go beyond the
lowest-order perturbative results of this work .

Now, we compute the backscattered current when the impurity is switched on at a finite time $t_{0}$:

\begin{equation} I_{bs}(t, t_{0}) = \frac{\Theta(\tau)\Gamma[2 K] \tau ^{2 - 2 K}}{\Gamma[K]\Gamma[2 - K]} \, \, _{1}F_{2}[ 1 - K ;  3/2 , 2 - K ; - (\tau/2)^{2}]\,I_{bs}(t, -\infty) ,\label{current}\end{equation}
where $_{1}F_{2}$  is the generalized hypergeometric function, and
$\tau = \frac{e V (t - t_{0})}{\hbar}$ is a dimensionless scaling
parameter so that $\tau \gg 1 (\ll 1 )$ represents a large (short)
time elapsed since the time $t_{0}$. This expression is the first
non-trivial result of this work. We have obtained an analytical
expression for the backscattered current taking into account the
effect of the nonadiabatic switching of the barrier. In the long
time regime this current approaches the stationary value $I_{bs}(t,
-\infty)$.

In order to perform a quantitative analysis of the transient process
which is now accessible due to the introduction of an abrupt
triggering mechanism, we found useful to evaluate the relative
change between currents turned on at times $t_{0}$ and $- \infty$,
$f(\tau)= \frac{I_{bs}(t, t_{0})}{I_{bs}(t, -\infty)} - 1$. This is
a damped oscillatory function of $\tau$ with period $2 \pi$ and
relative maxima in $\tau = (2 n + 1) \pi$ with $n$ natural. As
expected, $f(\tau)$ goes to zero when $\tau \rightarrow \infty$,
i.e, when the current is the one corresponding to a simple static
impurity acting at all times. At this point, in order to have a more
intuitive picture of the transient process, we define a reference
value of $\tau$, $\tau_{R}$ that enables us to identify time scales
for which the transitory stage is relevant or not. To this end it
seems natural to examine the way in which the relative maxima of
$f(\tau)$ decrease as $\tau$ increases. Since the first maximum is
located at $\tau=\pi$, it is a very good approximation to use the
following asymptotic ($\tau \gg 1$) expression for $f(\tau)$:

\begin{equation}  f(\tau)\approx  A(\tau,K)\,\cos[\tau + \pi] \label{g},\end{equation}
where $A(\tau,K)=  \frac{ \tau ^{- 2 K}}{\cos[\pi K] \Gamma[1 - 2
K]}$  is the envelope of a damped harmonic oscillation. Since the
change in the backscattered current due to the sudden switching
behaves as $\tau ^{- 2 K}$, we conclude that the relaxation of the
system is faster for small electron-electron interaction (For the
sake of clarity, let us stress that this relaxation characterizes
the transition of the backscattering current between two
off-equilibrium regimes). We thus find an explicit connection
between electron interactions and the switching time of the
externally controlled barrier: the stronger the correlations, the
longer the persistence of the non adiabatic effect. Formula
(\ref{g}) provides a direct way of defining a dimensionless
relaxation parameter $\tau_{R}$, such that for $\tau_{R} < 1 (> 1)$
the switching process is relevant (irrelevant). We define $\tau_{R}$
as the value of $\tau$ such that $A(\tau,K) = 1/r$ ($r>1)$, which
means that for $\tau > \tau_{R}$ the value of $I_{bs}(t, t_{0})$
differs in less than $100/r$ percent from the current obtained when
the switching process is not taken into account. To be definite, in
the following we set $r=10$. We then get
\begin{equation} \tau_{R} = \left(\frac{10}{\cos[\pi K]\Gamma[1 - 2 K]}\right)^{\frac{1}{2 K}}. \end{equation}

\begin{figure}\begin{center}
\includegraphics{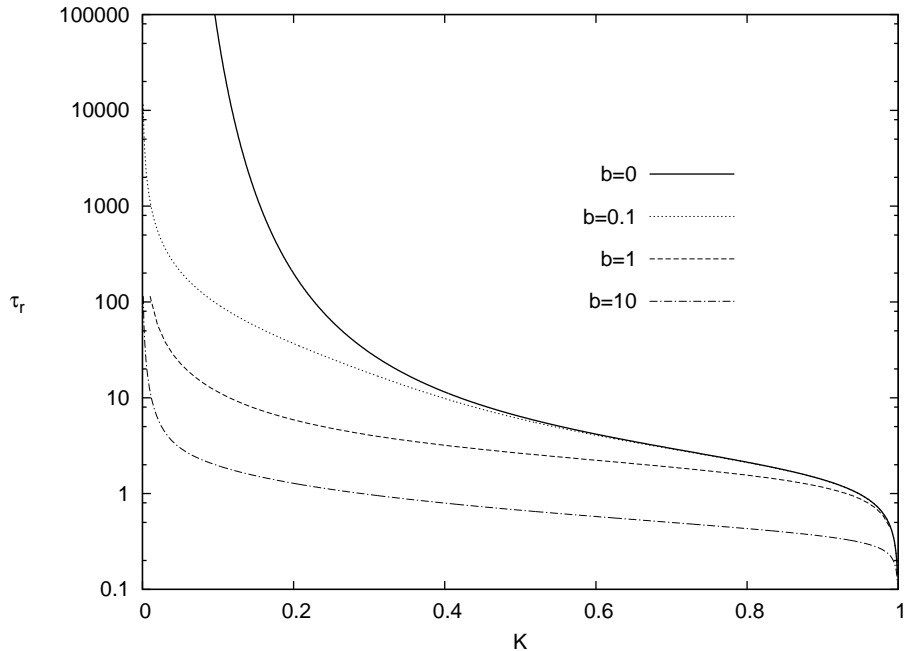}
\caption{\label{fig1ne} Dimensionless relaxation parameter
$\tau_{R}$ as function of $K$ and for  different values of $b$.}
\end{center}\end{figure}

The solid line in Figure \ref{fig1ne} shows the behavior of
$\tau_{R}$ as function of $K$ at zero temperature.  We observe that
$\tau_{R}$ grows as $10^{1/(2K)}$ for strong electron-electron
interactions ($K\rightarrow0$). For weak interactions
($K\rightarrow1$) the effect of the sudden switching is negligible,
$\tau_{R}$ goes to zero as $(1-K)^{1/2}$.

At this point it is useful to recall that the measurable time in
this process is the difference $t-t_0$. The time interval that it
takes the backscattered current to reach a value that differs in
exactly 10 percent from the steady current is $t_{R}-t_0 =
\frac{\hbar \, \tau_{R}}{e V}$. Using the results shown in Fig. 1
for the zero temperature case, which relate values of $K$ with
values of $\tau_{R}$, we can estimate $t_{R}-t_0$. Bias voltages
usually applied in experiments with nanodevices range from $\mu V's$
to $m V's$. On the other hand, typical interactions take values in
the range $.25<K<.75$ \cite{spin-charge}. With these data we obtain
$10^{-11}s <t_{R}-t_0 <10^{-7}s$, which are time intervals that
could be measured with recently developed pump-probe techniques
which reach femtosecond-attosecond time resolutions
\cite{femtosecondExp}.

The relationship between the relaxation process and the strength of
Coulombian electron correlations revealed in our analysis might
provide an alternative way to determine the Luttinger parameter $K$
through time measurements. The total current after the switching as
a function of time is $I(t) = I(\infty) + C t^{-2K}
\cos[{\Omega\,t}]$, where $\Omega=eV/\hbar$, $C$ is a constant and
$I(\infty)$ is the stationary current for a static barrier.  A
determination of the current as function of time with a temporal
resolution smaller than $\frac{2 \pi \hbar}{e V}$ is a direct method
to obtain the exponent of the temporal decay, and then, the $K$
value of the quantum wire. We emphasize that this proposed method is
performed at constant source-drain voltage.

In Figure \ref{fig3} we illustrate the experimental determination of
$K$ according to the ideas mentioned above. The solid line
represents the measurement of the total current $I(t)$ after the
switching of the impurity. All parameters involved in this plot (the
voltage and the strength of the barrier) are within the range of
experimental reach. The temporal resolution corresponding to this
hypothetical measurement is of the order of the picosecond, which is
accessible with the already mentioned state of the art pump-probe
techniques. The next step is to determine the envelope of the curve,
given by the dashed line in the figure. This envelope can be fitted
with the $t^{- 2 K}$ law in order to find $K$. This procedure, which
we depict here for $T=0$ can be extended for finite temperatures, as
we will show in the next section. The power law $t^{- 2 K}$ is a
very good approximation for $T\ll\frac{e V}{\pi k_B}$. For
source-drain voltages of the order of mV's, which we used in the
case of Figure \ref{fig3}, this corresponds to $T\ll 10^{\circ}K$.

\section{Results at finite temperature}

Taking into account that experiments are performed at very low but
finite temperatures, in this section we show how our results are
affected by thermal effects. Using expression (\ref{tres}) for
vacuum expectation values of exponentials of bosonic fields, we can
extend the results of Section III in a straightforward way. The
backscattered current with the impurity switched on at time $-
\infty$ (static impurity) becomes:

\begin{equation}
I_{bs}(t, -\infty) = \frac{g_{B}^{2} e  \Lambda ^{2 K - 2}}{2 \pi^{2} \hbar^{2} v ^{2 K}\Gamma[2 K]}(2 \pi k_{b} T/\hbar)^{2 K -1}\Gamma[K - \frac{i}{2 b}]\Gamma[K + \frac{i}{2 b}]\sinh[\frac{\pi}{2 b}],
\end{equation}
where $b = \frac{\pi k_{b} T }{e V}$ is a dimensionless parameter that characterizes the scale regime in temperature of the system: $b \gg 1$ ($\ll 1$) is the  high (low)-temperature regime, for fixed voltage. The perturbative expansion is valid for low temperature
when $\frac{g_{B}}{\hbar v} (\frac{\Lambda e V}{ \hbar v})^{K - 1} \ll 1$ and for high temperature
when $\frac{g_{B}}{\hbar v} (\frac{\Lambda k_{b} T}{ \hbar v})^{K - 1} \ll 1$.

The backscattered current when the impurity is switched on at a finite time $t_{0}$ is:

\begin{equation}\label{general} I_{bs}(t, t_{0}) = \left\{1 +  \frac{i \Theta(\tau)\Gamma[2 K] \sin[\pi K] (B[\exp[2 b \tau], \frac{- i}{2 b}, 1 - 2 K] - \ast) }{\Gamma[\frac{- i}{2 b} - K]\Gamma[\frac{i}{2 b} - K]\sinh[\frac{\pi}{2 b}]} \right\}I_{bs}(t,-\infty) ,\end{equation}
where $B$ is the Incomplete Beta function and $\ast$ indicates the
conjugate of the precedent term. This is the analytical
generalization of (\ref{current}) at finite temperature. As in the
former case, the relative change between currents turned on at times
$t_{0}$ and $- \infty$ is a damped oscillatory function of $\tau$
with the same period and position of relative maxima as before.
Using the asymptotic approximation, we obtain the following
expression for the envelope of the damped oscillation:
\begin{equation}
A(\tau,K,b) \approx  \frac{4 b \Gamma[2 K] \sin[\pi K] }{(2 \sinh[b \tau] )^{2 K}\Gamma[K - \frac{i}{2 b}]\Gamma[K + \frac{i}{2 b}]\sinh[\frac{\pi}{2 b}]},
\end{equation}
Thus, the change in the backscattered current due to the sudden
switching at finite temperature behaves as $\sinh [ b \tau] ^{- 2 K}
= \sinh [ \pi k_{b} T (t - t_{0})/\hbar ]^{- 2 K}$ .  For high
temperatures,  ($b \gg 1$), the decay is faster and becomes
exponential: $A(\tau,K,b) \approx \frac{8 b^{2} \Gamma[2 K] \sin[\pi
K] \exp[ - 2 K b \tau]}{\Gamma[K]^{2}}$.

Experimentally, the high temperature regime can be accessed for low
voltages. Temperatures involved in the crossover regime ($b \simeq
1$) are $T \simeq 10^{\circ} K$ and $T\simeq 0.01^{\circ} K$, for
applied voltages of order $1 mV$ and $1 \mu V$, respectively. In
general, if we know the temperature of the system, a method
completely analogous to the one described at the end of the previous
section can be established to obtain the parameter $K$.  The total
current after the switching is $I(t) = I(\infty) + C \sinh [ \pi
k_{b} T (t - t_{0})/\hbar ]^{- 2 K} \cos[{e V t/\hbar}]$. Thus, the
value of $K$ can be obtained by measuring the current as a function
of time at fixed voltage.

As in the case of zero temperature, we can define a dimensionless
relaxation parameter $\tau_{R}$ in the same fashion; i.e. by
determining the value of $\tau$ such that $A(\tau,K, b) = 0.1$.  We
obtain for $\tau_{R}$ the following general expression as function
of temperature and electron-electron interaction strength:

\begin{equation}\label{relax} \tau_{R} = \frac{1}{2 b} \arcsinh \left[\left(\frac{40 b \Gamma[2 K] \sin[\pi K] }{\Gamma[K - \frac{i}{2 b}]\Gamma[K + \frac{i}{2 b}]\sinh[\frac{\pi}{2 b}]}\right)^{\frac{1}{2 K}}\right]. \end{equation}

Figure \ref{fig1ne} shows the dimensionless relaxation parameter
$\tau_{R}$ of the system as function of $K$, for different values of
temperature. We observe that the distortion in the backscattered
current caused by the transient process is more important for high
electron-electron interactions ($K \ll 1$) and in the regime of zero
or low temperature ($b \ll 1$), where $\tau_{R} \gg 1$. In this case
the relaxation is very slow and the sudden switching changes
significantly the value of the backscattered current for a long
time. If $b$ and $K$ grow, this effect tends to disappear.  For high
temperatures, the switching effect becomes irrelevant.

\section{Conclusions}
We have theoretically analyzed the consequences of a sudden switch
of a tunnel barrier on transport properties, in a long and clean LL.
In particular we studied the behavior of the backscattered current
$I_{bs}$ in a system subjected to a bias voltage V. The
backscattering time-dependent impurity is assumed to be point-like
and weak. Under these conditions, using bosonization and our result
(\ref{tres}) for the expectation value of exponentials of bosonic
operators, we obtained an analytical expression for $I_{bs}$ as a
function of $\tau = \frac{e V (t - t_{0})}{\hbar}$ ($t_0$ is the
instant at which the barrier is switched on), $b = \frac{\pi k_{b} T
}{e V}$ and the Luttinger parameter $K$ (See formula
(\ref{general})). At long times $t\gg t_0$, $I_{bs}(t,t_0)$ reaches
a steady-state value that coincides with the value of the current
corresponding to a static impurity (a barrier switched on at an
infinitely remote instant in the past) $I_{bs}(t,-\infty)$. By
carefully examining the way $I_{bs}(t,t_0)$ approaches
$I_{bs}(t,-\infty)$ as time goes by, we intuitively characterized
the transient process defining a dimensionless relaxation parameter
$\tau_R$. Employing an asymptotic expression for the exact solution
(\ref{general}), we obtained a simple expression for $\tau_R$ as
function of $b$ and $K$ (See formula (\ref{relax}) and Fig. 2). From
this result one concludes that, for fixed bias, $\tau_R$ grows with
decreasing $K$, meaning that transient effects on the current are
expected to last longer when the electron correlation is higher. In
fact, $\tau_R \rightarrow 0$ for a free system ($K \rightarrow 1$).
One can also see that the transitory effect is more pronounced for
low temperatures ($b\ll1$).

Finally, we would like to stress that our study could have some
interesting experimental applications. Indeed, from our main result
(\ref{general}), and taking into account that the total current
after the switching is $I(t)= I_0 - I_{bs}(t,t_0)$, we obtained
$I(t) = I(\infty) + C \sinh [ \pi k_{b} T (t - t_{0})/\hbar ]^{- 2
K} \cos[{e V t/\hbar}]$. This means that, at fixed temperature and
bias voltage, the value of $K$ of a 1D structure could be determined
by performing measurements of the total current as function of time,
after turning on a tunnel barrier through a localized gate voltage,
following, for instance, the techniques of reference \cite{biercuk}.

\vspace{1cm} This work was partially supported by Universidad
Nacional de La Plata (Argentina) and Consejo Nacional de
Investigaciones Cient\' ificas y T\'ecnicas, CONICET (Argentina).
The authors are grateful to M. Di Ventra for a helpful e-mail
exchange, and A. Iucci for useful comments.

\end{document}